\documentclass{an}
\usepackage[sort]{natbib}
\usepackage{graphicx}
\usepackage{times}
\usepackage{fancyhdr}
\def\aap{\rm{A\&A}}                

\begin{document}

\title{An autonomous adaptive scheduling agent for period searching}

\author{Eric S.~Saunders\inst{1,2}\fnmsep
\thanks{Corresponding author:\email{saunders@astro.ex.ac.uk}\newline}
        \and Tim Naylor\inst{1}
        \and Alasdair Allan \inst{1}
        }
\institute{School of Physics, University of Exeter, Stocker Road, EX4 4QL
           \and
           Las Cumbres Observatory, 6740 Cortona Dr.,
           Suite 102, Santa Barbara, CA 93117}

\date{Received; accepted; published online}

\abstract{ 
   We describe the design and implementation of an autonomous adaptive software
   agent that addresses the practical problem of observing undersampled,
   periodic, time-varying phenomena using a network of HTN-compliant robotic
   telescopes. The algorithm governing the behaviour of the agent uses an
   optimal geometric sampling technique to cover the period range of interest,
   but additionally implements proactive behaviour that maximises the optimality
   of the dataset in the face of an uncertain and changing operating
   environment.
}

\maketitle

\section{Introduction} \label{Section:Intro}


The eSTAR project \citep{allan04_estar_spie, allan07_estar_htn_iii} is an
agent-based software system that aims to establish an intelligent robotic
telescope network for efficient and automated observing. The system comprises
\emph{user agents} that run an observing programme on behalf of an astronomer,
and \emph{embedded agents} that provide the interface to observational resources
such as telescopes. One role of eSTAR is to provide the software
infrastructure for Robonet-1.0 \citep[e.g.][]{bode04robonet_canonical, mottram06_high_level_robonet}, a network
of three autonomous 2-m robotic telescopes strategically located across the
globe. eSTAR implements the HTN\footnote{Heterogeneous Telescope Networks}
protocol \citep{allan06_htn_protocol_standard} using the XML dialect RTML
\citep{pennypacker02_rtml2.1, hessman06_rtml3}. The Robonet-1.0 network thus
provides a full reference implementation of an HTN-compliant environment.


This paper describes the design and implementation of a user agent that applies
adaptive scheduling to optimise an observing run. Adaptive scheduling was
described in an earlier paper \citep{saunders06_agent_metrics} as ``techniques... to allow an autonomous software
entity to calculate an idealised observing strategy, while remaining able to
respond flexibly to scheduling constraints outside of its immediate
control.'' In particular, the adaptive
scheduling agent (ASA) implements the optimal geometric sampling strategy for
period searching described in \citet{saunders06_optimal_placement} to address
the practical problem of observing undersampled, periodic, time-varying
phenomena.


An example of the type of time-domain astronomy that this agent was designed to
perform is the surveying of variability in star-forming regions. Especially in the youngest
clusters, a large fraction of the cluster members are T-Tauri stars that can
exhibit significant variability due to rotational modulation of features at the
stellar surface. However the range of periods among cluster members is large,
ranging from a few hours to many days. Additionally, aliasing, particularly as a
consequence of diurnal sampling, is usually a problem for data obtained in the
``classical paradigm'', where an on-site observer performs observations using a
single telescope. A robotic network spread across longitude can break this aliasing pattern, but
datasets are typically undersampled with respect to shorter periods in the range
of interest. Given a limited number of observations, the optimal geometric
sampling technique determines the best time to make individual observations in
order to achieve similar sensitivity to periods across the full range of
interest. Of particular importance is the property that as long as the set of
gaps between observations remains unchanged, observations may be arbitrarily
reordered \citep{saunders06_optimal_placement}. This property is fully exploited
by the adaptive algorithm.


Unfortunately, observations are not guaranteed. In the dispatch model of
telescope scheduling, users request observations but it is up to the telescope
scheduler to decide whether such requests are honoured
\citep{fraser06_robonet_scheduling}. Observations may still fail even when
submitted to an entirely cooperative scheduler, due to the possibility of
telescope downtime or inclement weather. Thus any system seeking to reliably
implement an optimal geometric sampling technique must in practice deal with
such observation uncertainty.

\section{Design principles}


The environment in which the adaptive scheduling agent operates is inherently
uncertain. This fact was the fundamental design constraint from which the
desirable attributes of the agent were derived. Four principal ideas drove the
design.

\begin{itemize}

\item{\bf Robustness.} Building software that will operate reliably over a
network is complicated. Therefore a guiding principle was to keep the algorithm
as simple as possible, to minimise the possibility of unforseen interactions
between system components. With a distributed system it must be assumed that any
remote part of the system can fail at any time, either temporarily or
permanently. The algorithm needs to be able to make a reasonable attempt to
continue under such circumstances.

\item{\bf Cynicism.} Although all the agents in the eSTAR system are in
principle trustworthy, in practice it is much safer to assume nothing binding
about events or entities external to the main code. The benevolence assumption,
namely that agents may act for the greater good of the system, at their own
expense, is not valid here. This is because the telescope schedulers, and by
proxy the actions of the embedded agents, are not under the control of eSTAR.
The goal of the embedded agents is to optimise the schedule in some internally satisfactory way,
which may or may not coincide with the goals of the user agent. This is most
important with respect to the scoring information returned by the observing
nodes: since there is no penalty to the embedded agent for providing an
inaccurate score, there is no compelling reason for the user agent to trust that
value. Even if there is reason to believe that the node is acting in good faith, we still have no
\emph{a priori} idea of the accuracy of the information being supplied.
Therefore external information should be considered, but not relied on, and in
general the agent needs to adequately handle the worst case scenario (being
supplied with false information).

\item{\bf Stability.} A poor algorithm would require chains of specific events
in order to be successful. If the future is highly uncertain, then an algorithm
that is reliant on that future to succeed is risky and fragile. Ideally, the
performance of the algorithm should degrade gracefully as the environment
becomes more hostile, so that it continues to make a best effort to succeed. In
practical terms, this means that partial runs need to be optimal, i.e. that
whatever the agent has achieved to date needs to be the best set of observations
that it could make under the circumstances, since we cannot guarantee future
observations at the temporal positions we specify.

\item{\bf Adaptability.} Finally, the agent needs to constantly evaluate the
current state of the run and the current observing conditions, and alter its
behaviour as necessary. Some simple examples illustrate this point. The agent is
aiming to acquire a set of observations that are spaced by both small and large
gaps of particular sizes. If the agent has acquired many small gaps, how should
it alter its behaviour to try to achieve larger gaps? What should it do in the
opposite case? Indeed, how should it deal with the general problem of a set of
gaps that are not quite correct? Another plausible scenario is that of a
telescope that consistently provides low scores, but still manages to complete
observations successfully. How should that be handled? The agent programming
approach explicitly assumes uncertainty and change, and defines a feedback cycle
to interact with the environment. How to successfully implement useful adaptive
behaviour is of critical importance to the success of this approach.

\end{itemize}

\section{Implementation}

\begin{figure}
   \centering 
   \includegraphics[width=\columnwidth]{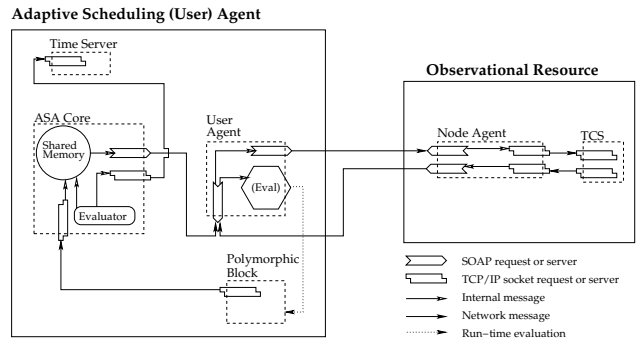}
   
   \caption{
   Architecture of the adaptive scheduling agent. The multithreaded ASA core
   implements the adaptive evaluation engine, a SOAP submission mechanism for
   making observation requests, and a TCP/IP socket listener for receiving
   observation feedback information. Date/time information is abstracted by a
   time server, which can be accelerated to allow fast simulations to be
   performed. Actual RTML document construction and HTN protocol negotiations
   are handled by the user agent, a web service that implements a SOAP server
   for submitting observations and receiving asynchronous messages from the
   embedded agent. Asynchronous messages trigger a run-time code evaluation
   based on the observation type, allowing arbitrary response behaviour to be
   defined for different observing programmes. In this case, the polymorphic
   block strips pertinent information (primarily the observation timestamp) from
   the incoming message and pipes it to the dedicated socket listener in the ASA
   core. In this way the agent receives observation feedback as the run
   progresses, allowing direct optimisation of the future observing strategy.
   \label{fig:asa_detail}}
\end{figure}

Figure \ref{fig:asa_detail} shows the detailed architecture of the adaptive
scheduling agent. The core decision-making algorithm is a multi-threaded Perl
process that determines when to observe, translates between abstract theoretical
timestamps and real dates and times, and monitors the progress of the observing
run.


Observation requests are made by remote method invocation of the user agent web
service running on the same machine. Following the HTN protocol
\citep{allan06_htn_protocol_standard}, the user agent makes scoring requests of
each telescope (by way of the embedded agent at each site), and then picks the
highest non-zero score. A ``request'' document is then submitted. Under normal
conditions, a ``confirmation'' document is returned. This indicates that the
observation request has been accepted by the telescope and is now queued.


Eventually, the observation request is resolved. Although multiple exposure
observation requests are possible, the ASA always requests single observations,
as this maximises the amount of scheduling control retained by the agent. Either
the observation was successful, and an ``observation'' document is returned, or
the observation did not succeed, and a message of type ``fail'' is returned. The
return document is received by the user agent web service, and scanned for
observation type. The agent compares the type to the set of types it knows how
to deal with, and if there is a match, the code for that type is dynamically
loaded and executed. This plug-in architecture allows custom actions to occur
for different types of observation, allowing a single instance of the user agent
web service to handle RTML marshalling for many distinct observing programmes.


The algorithmic block for the ASA extracts the critical pieces of information
from the incoming RTML document. These include the requested start time, and if
the observation was successful, the actual start time of the observation,
gleaned from the FITS header included in the ``observation'' document. Any error
messages are also picked up here. This subset of key information is then
transmitted by a simple socket connection to the listener thread of the ASA
core code. In this way the loop is closed.

\section{Algorithm}

\subsection{Optimality}


If an agent could place observations at any time and be guaranteed of success,
then the choice of observations is clear: they should be placed with the gap
spacings indicated by the optimal sampling. However in the normal operating
environment many observations can fail. When an observation eventually does
succeed, the gap between that observation and the last successful observation is
unlikely to be of an ideal length --- but it could be close.


What the agent requires is some unambiguous way to determine how well its
completed spacings compare to the optimal set of gaps. It is not possible to
simply compare the two sets of gaps and ``tick off'' perfect gaps as they are
obtained, because even a ``correct'' observation is not precisely located.
Telescope overheads mean that in practice an acceptable window for the
observation must be provided, and the observation can take place anywhere within
that window. Some sort of fuzzy criterion could be used, but this must be
explicitly defined and is somewhat arbitrary.


\begin{figure}
   \centering 
   \includegraphics[width=\columnwidth]{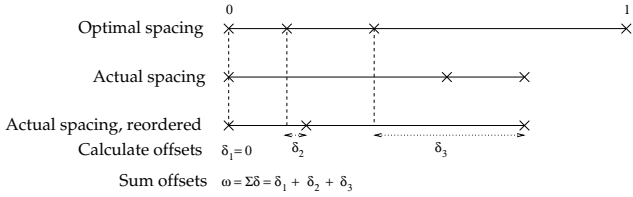}
   
   \caption{Calculating the optimality, $w$, of a set of observations. The
   optimal and actual sets of gaps are sorted, smallest to largest, and
   each gap compared. The sum of offsets is $w$.
    \label{fig:optimality}} \end{figure}


The \emph{optimality criterion}, $w$, is defined by the following simple
series of operations.

\begin{enumerate}
\item{Order the set of optimal gaps, from smallest to largest.}
\item{Order the set of obtained gaps, from smallest to largest.}
\item{For each obtained gap, find the offset of that gap from the optimal gap at
that position in the set.}
\item{The overall optimality, $w$ is the sum of these offsets.}
\end{enumerate}


Figure \ref{fig:optimality} illustrates this process. The optimal gap set is
expressed in theoretical time, i.e. so that the run begins at timestamp 0 and
the final observation occurs at timestamp 1. The obtained gaps are scaled to the
same units. If the set of actual timestamps is in perfect agreement with the
optimal set, then the value of the optimality metric is 0. There is no upper
limit on the value since the length of the actual run can exceed that of the
optimal sequence.


Note that the simplicity of this approach is only possible because of the
reordering property of the optimal sequence. This allows the gaps to be compared
in order of size, regardless of the actual observed ordering.

\subsection{Choosing the next observation}


The optimality function allows future observation timestamps to be compared. The
agent seeks to minimise the value of the optimality, which will increase
monotonically as more observations accrue. The question is, given the existing
set of observations, what new timestamp will achieve this? Since time is
continuous, in principle there are an infinite number of possible futures to
choose between. Once again however it is the ability to reorder optimal
observations that allows some elegant simplifications to be made. Firstly, if
all reorderings of a given optimal series are considered equally optimal, then
the set is degenerate with respect to reordering: for any set of gaps we need
only consider one possible reordering. The most straightforward is to place the
gaps in order of size. Secondly, it is apparent that any new observation should
be placed in such a way as to exactly achieve one of the optimal gaps, since
anything else would immediately increase the value of $w$ for no gain. This
insight drastically reduces the search space of possible timestamps.


An important side-effect of the definition of the optimality is that the agent
usually prefers to acquire shorter gaps first. This behaviour emerges because
the optimality is calculated by comparing optimal gaps to actual acquired gaps.
The  magnitude of the increase in $w$ accrued from any new observation is
directly proportional to the distance of that gap from its comparison gap.
Because we choose to order the comparison gaps from shortest to longest, this
means that large gaps early on are penalised by relatively large increases in
$w$. This is an important feature. It makes good practical sense to go for the
shortest useful gap at any given time, because the agent must wait until that
timestamp before it can discover whether the observation actually happened (and
hence the required gap collected).


It should be noted, however, that it is not sufficient to always seek the shortest
``uncollected'' gap, observation after observation. This would indeed be the best
strategy if observations were always guaranteed to succeed, but in practice two
types of forced deviations from optimality can be expected to occur. The first
is that some timestamps are known prior to observation to be inaccessible, and
hence are guaranteed to fail (e.g. because it is day, the target is not visible,
the target is too close to the moon etc.). This information is passed back to
the agent in the form of a score of 0. The second is that some fraction of
observations that were expected to succeed will in fact fail ``on the night,'' due
to uncertainties in the observing process itself (weather, telescope load,
etc.).


These failures force the creation of sub-optimal gaps in the observing pattern.
The specific shape of this observed pattern drives the choice of the next gap.
This is because sub-optimality is defined as a continuous spectrum --- a gap of
a similar size to an optimal gap is ``more optimal'' than a gap of a very
different size. The effect of failed observations is to increase the average
size of gaps in the observed set. Although these larger gaps were not planned,
they can nevertheless be useful, since they will be nearer in size to the larger
gaps in the optimal set. It may therefore make sense for the agent to aim for a
smaller gap, because this will shunt the larger gaps along in the optimality
comparison, improving the overall optimality of the series. In other situations
however, the agent may choose a slightly longer gap, even if it has not achieved
all the shorter gaps preceding it, because too many short gaps will worsen the
optimality of the longer end of the run. This is a direct consequence of finding
a better value of $w$ for a longer gap.


It is worth noting that at no point are the data themselves examined. This is by
design; observing the Fourier transform of the data during the experiment, for
example, and changing the sampling behaviour based on the results leads to a
complex quagmire of interactions between the observer and the observations. It
is possible, for example, to home in on a spurious signal in the periodogram,
and to choose a sampling that selectively reinforces that signal. The guiding
principle of simplicity suggests that such complexities should be avoided.

\section{Testing and results}

The agent was extensively tested in a virtual telescope environment before being
deployed on the eSTAR network, where it acquired observations of a variable star
with a well-known period as a proof-of-concept. This environment was designed to
provide a virtual telescope network that would replicate as closely as possible
the actual operating conditions under which the agent was to run. Each virtual
telescope instance can be ``sited'' at a different virtual location, allowing the
calculation of sunrise and sunset times, and the rise and set times of arbitrary
points on the celestial sphere. To make the environment more challenging,
non-zero score replies have random values between 0 and 1, and have no
correlation with the likelihood that the observation will in fact be successful.

The passage of time in the simulation is regulated by the \emph{time server}.
This is a simple standalone process that provides the current simulation time
via a socket interface. Internally, the time server calculates the current
simulation time by scaling the amount of real time elapsed since instantiation
by an acceleration factor, provided at startup. Since all agents use the time
server as the canonical source for timestamps and timing calculations, the time
server allows the simulation to be run many times faster than the real world,
enabling full simulations of a likely observing run to complete in a reasonable
timeframe. Setting the acceleration factor to 1 allows the timeserver to run at
normal speed, and therefore provides transparent access to the real-world time.

The probability of observation success is specified at startup for each virtual
telescope, and a random number generator used to determine the success or
failure of each observation at run-time. If successful, a fake FITS header is
generated, with the correct observation timestamp placed within the header. This
header is encapsulated inside the RTML message of type ``observation'' that is
returned. Otherwise, a ``fail'' message with no data product is returned. Using
the existing user agent and node agent codebase meant that at the protocol level
the interaction between eSTAR components in the simulator was identical to that
in the real world, allowing most aspects of document exchange to be tested
before live deployment.

One limitation of the simulation environment was that fail rates at the virtual
telescopes were not correlated, that is, periods of failure were not consecutive
in time. This was not implemented because of the complexity of choosing
realistic correlation behaviour. Even without such behaviour, the most critical
aspects of the agent were able to be adequately tested. Nevertheless, this is an
obvious way in which the virtual telescope environment could be improved.

The purpose of the tests performed in simulation was to identify bugs at several
levels, and to evaluate the performance of the algorithm and iterate
improvements in a tight feedback cycle. This included analysis at the
message-passing level for conformity with the HTN protocol specification. The
behaviour of the implemented algorithm was carefully compared with the design
statement. A number of discrepancies were found and corrected. Most importantly,
the performance of the algorithm under pseudo-realistic conditions allowed a
number of unforeseen corner cases to be identified and correctly handled.

The scheduling algorithm was tested in a number of simulated observing runs, but
the discussion of a single experiment is sufficient to illustrate the general
performance characteristics. In this simulation a 10 day, 60 observation run was
attempted. An acceleration factor of 50, applied to the time server, allowed a
single 30 minute observing window to complete in 36\,s. The start of the run was
set to a date of 27/04/07, at which time the target was observable approximately
2/3 of the time from at least one of the three Robonet-1.0 virtual telescopes.
In this simulation, in addition to the non-observable blocks (which returned a
score of 0 to the agent), the virtual telescopes were set to randomly fail
observations at run-time with 50\% probability (an arbitrary fraction that
seemed reasonable).

The experiment was halted after 10 days had elapsed in simulation time. It was
found that of all the observations submitted to the network, 40 had succeeded
(i.e., been accepted to the queue, and then been declared successful at some
point in the specified observing window), while 50 of the queued observations
had failed (queued but not observed). All the successful observations were found
to have close-to-optimal timestamps, subject to the observability and success
constraints.

The results indicated crudely that if the real network exhibited a similar fail
rate, then a ballpark figure of approximately 2/3 of the observations to be
completed successfully was a reasonable expectation by the 10th continuous day
of observing, and additionally, that the spacings between individual
observations were likely to be near-optimal.

The analysis of the results of the subsequent observing run are outside the
scope of this work, but will be presented in detail in a forthcoming paper
\citep{saunders07_estar_asa_results}. For a detailed discussion of optimal
geometric sampling and the adaptive scheduling agent, including results, see
\citet{saunders07_thesis}.

\section{Conclusions}

We have built an adaptive scheduling agent that implements the optimal geometric
sampling strategy for period searching described by
\citet{saunders06_optimal_placement} in the context of an HTN-compliant network
of telescopes. The algorithm automatically penalises the agent for choosing too
many short gaps, but is similarly hostile to the chasing of large gaps when
there are shorter, more valuable alternatives. In this way a balance point is
determined that unambiguously defines the best option for the agent at any point
in the run. The deceptively simple rule sequence for finding the optimality
implicitly utilises the reordering property of optimal sequences, maximises the
agent's cautiousness in the face of uncertainty, and provides a computationally
cheap and scalable way to unambiguously calculate the best choice of action at
any point. It is also stable --- a snapshot of the run at any point in time is
optimal for that subset of points, and minimises the effects of observation
failure on the sequence. In the case of perfect observing behaviour (no
failures), the optimal sequence is naturally recovered. Importantly, the degree
of failure exhibited by the network can also change dynamically without
adversely affecting the algorithm, because it makes no assumptions about the
stability of the network, and makes no attempt to model the behaviour of the
telescopes on which it relies.

\acknowledgements 
   ESS was funded through a PPARC e-Science studentship. This work is part of
   the eSTAR project, which supports AA and which is jointly funded by the DTI,
   PPARC and EPSRC.


\end{document}